\documentclass[a4paper,fleqn]{cas-dc}

\usepackage[numbers,sort&compress]{natbib}
\usepackage{subfigure}
\usepackage{bm}
\usepackage{ulem}
\usepackage{multirow}
\usepackage{array}

\begin{document}
\let\WriteBookmarks\relax
\def\floatpagepagefraction{1}
\def\textpagefraction{.001}
\shortauthors{Qian Wu et~al.}

\title [mode = title]{Probing the existence of $\eta^3$He mesic nucleus with a few-body approach}                

\author[1,2]{Qian Wu}
\ead{qwu@nju.edu.cn} 

\author[2]{Gang Xie}

\author[2,3]{Xurong Chen}

\address[1]{School of Physics, Nanjing University, Nanjing 21000, China}
\address[2]{Institute of Modern Physics, Chinese Academy of Sciences, Lanzhou 730000, China}
\address[3]{School of Nuclear Science and Technology, University of Chinese Academy of Sciences, Beijing 100049, China}


\begin{abstract}
Motivated by the two recent observations in the WASA-at-COSY detector, we investigate the $\eta^3$He nucleus with the $\eta NNN$ few body method.
We construct the effective $s$-wave energy dependent $\eta N$ potential which reproduce the $\eta N$ subthreshold scattering amplitude
in the 2005 Green-Wycech model. It gives the $\eta$ separation energy and decay width of 0.19 MeV and 1.71 MeV, respectively.
We also construct various sets of effective $s$-wave energy independent $\eta N$ potentials where the corresponding 
complex scattering lengths ($a$) are within the range given in most theoretical models. 
We obtain the bound $\eta^3$He nucleus with decay width of about 5 MeV when $a$ is (1.0 fm, 0.3 fm), and of about 10 MeV when $a$ is (1.0 fm, 0.5 fm).
\end{abstract}


\maketitle

\section{Introduction}\label{sec:1}
The $N^*(1535)$ resonance, which is close to the $\eta$-nucleon ($N$) threshold ($E_{\rm th}$=1487 MeV), results in
a strong attractive force between the $\eta$ meson and the nucleon.
It is first examined by Bhalerao and Liu in the pioneering work \cite{liu1985}, using
the $\eta N$-$\pi N$ coupled channels method. Soon it's verified in the dynamical calculation of the
$N^*(1535)$ $S_{11}$ resonance \cite{plb10}.
Since then, the $\eta N$ interaction has been
studied with several coupled-channel models which generate out a wide range of the real part of the $\eta N$ scattering length from 0.2 to 1.0 fm \cite{plb10,plb11,Green2005,Metag2017,Sibirtsev2001,Khreptak2023}.
At these works, the imaginary $\eta N$ scattering lengths are found to have a narrower range from 0.2 to 0.5 fm. 

Due to the attractive $\eta N$ interaction, it is possible to form $\eta$ mesic bound states in nuclei. The evidences that the $\eta$ mesic quasibound states may exist are given in Refs.~\cite{plb3,plb3-1,plb4,plb4-1}. 
Since then, various optical model calculations have been used for searching the $\eta$ nuclear bound state~\cite{npb5,npb6,npb7,npb8,npb9}.
In Ref.~\cite{xie2017}, Xie et al. obtained a 0.3 MeV's binding energy of the $\eta^3$He and a decay width around 3 MeV by evaluating the $pd\rightarrow\eta^3$He near-threshold reaction. A recent interpretation
of $\eta$ quasibound states constrained data in the photon and hadron induced reactions implies that
$\eta d$ is unbound, $\eta^3$He might be bound while $\eta^4$He is bound~\cite{Krusche2014}.

Besides the optical potential model calculations, there are also several few-body calculations concerning the $\eta NN$, 
$\eta NNN$ or $\eta NNN$ systems~\cite{gal2015plb,gal2017npb}.
With precise few-body stochastic variational method (SVM) and several energy dependent $s$-wave $\eta N$ interactions derived from several coupled channel models of the $N^*(1535)$ resonance,
Barnea et al. \cite{gal2017npb} reported a bound $\eta^3$He nucleus in the condition that the real part of the $\eta N$ scattering length should exceed 1.0 fm approximately which yields a few MeVs binding energy between $\eta$ and $^3$He. For the $\eta^4$He, they found that the real $\eta N$ scattering length should be at least 0.7 fm to form a bound nuclei. 
Similar conclusions were given with applying SVM and a pionless effective field theory (EFT) \cite{gal2017plb-pion}.
However, in Ref.~\cite{fix2017}, with solving the four and five body Alt–Grassberger–Sandhas equations,
it said neither $\eta^3$He and $\eta^4$He are bound when the $\eta N$ scattering length is $0.97\;{\rm fm}+0.27i$ fm.

On the experimental side, various experiments using photon, pion, proton or deuteron beams
have given signals of bound $\eta$ mesic nuclei~\cite{2020plb-2,2020plb-3,2020plb-4,2020plb-5,2020plb-8,2020plb-9} but none of them can conclude clear existence~\cite{2020plb-10}.
A very recent experimental campaign of searching the $\eta$ mesic nuclei was conducted by the WASA-at-COSY collaboration. 
The search for the $\eta^3$He bound state in the WASA-at-COSY detector were realizd through the $pd\rightarrow(^3\rm{He},\eta)\rightarrow(^3\rm{He},2\gamma)$ and $pd\rightarrow(^3{\rm He},\eta)\rightarrow(^3{\rm He},6\gamma)$ reactions. They reported
a $\eta^3$He bound nucleus with decay width $\Gamma$ above 20 MeV and binding energy $B_\eta$ between 0 and 15 MeV~\cite{Adlarson2019plb}. But this observation is within the range of the systematic error which doesn't allow one to make a definite conclusion.
The search in the $pd\rightarrow(^3{\rm He},\eta)\rightarrow(dp,\pi^0)$ reaction gives 13 to 24 nb for the bound $\eta^3$He giving the decay width between 5 and 50 MeV and the
$B_\eta$ between 0 and 40 MeV~\cite{Adlarson2019prc}.

Considering the uncertainty (large error bar) in the observation of the binding energy and decay width of the $\eta^3$He nucleus in the WASA-at-COSY detector, we investigate the $\eta^3$He system under various sets of effective $s$-wave $\eta N$ potentials with the few-body method, which allows us to have a certain range of the binding energy and the decay width.
We construct several sets of energy independent $\eta N$ potentials where the $\eta N$ scattering lengths 
are contained within the range of the values given in most theoretical literature. 
Besides, since the scattering amplitude given in most theoretical models declines rapidly as going deeper into the subthreshold energy region, a energy dependent $\eta N$ potential might be necessary. 
Thus in this work, we also apply the energy dependent $\eta N$ potential in the study of the $\eta^3$He nucleus.

Based on the above calculations, we will show the correspondences between the theoretical sets of $\eta N$ potentials and the  
binding energies (decay widths) of the $\eta^3$He nucleus within the framework of few-body system. 
Then, if people have more definite conclusions or more certain values from the
experimental side in the future, we will filter out the suitable $\eta N$ potential if it reproduces the experimental values.
This might help us have a better understanding of the $\eta N$ hadronic interaction and the $\eta$ mesic nuclei.

The paper is organized as follows: In Sec.~\ref{sec:21}, we introduce the method we used to solve the four body system.
Then in Sec.~\ref{sec:22}, we introduce the effective energy dependent $s$-wave $\eta N$ potentials which reproduce the scattering amplitude given by the GW model~\cite{Green2005}.
The energy independent $s$-wave $\eta N$ potentials giving the relations with the $\eta N$ scattering lengths are also given.
In Sec.~\ref{sec:31} and Sec.~\ref{sec:32}, the binding energy and the decay width of the $\eta^3$He nucleus are investigated respectively by using the energy dependent and the energy independent $\eta N$ potentials.
Sec.~\ref{sec:4} is devoted to summary.

\section{Methodology}
\subsection{Gaussian expansion method}
\label{sec:21}
\begin{figure}[htbp]
\setlength{\abovecaptionskip}{0.cm}
\setlength{\belowcaptionskip}{-0.cm}
\centering
\includegraphics[width=0.48\textwidth]{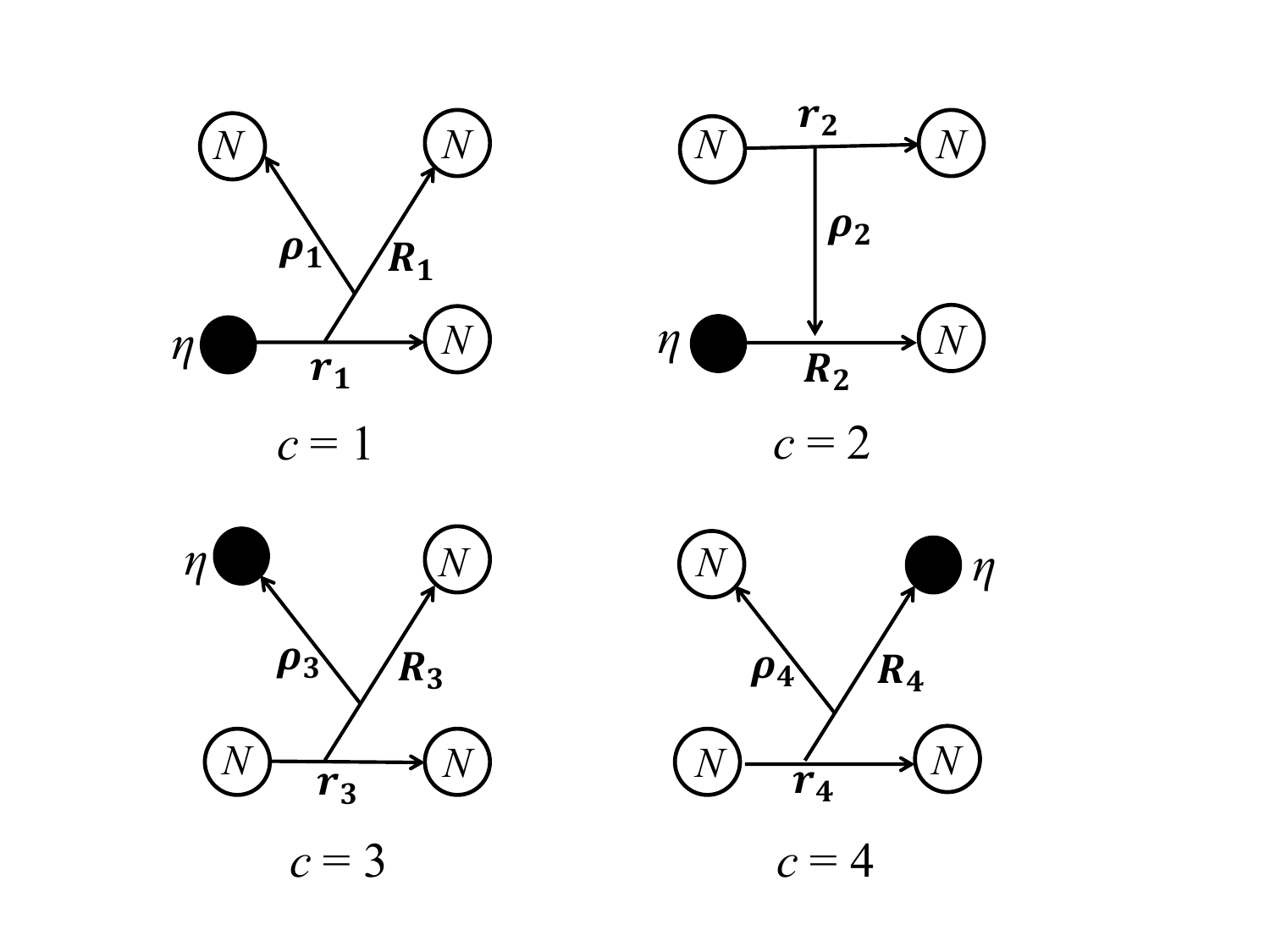}
\caption{Jacobi coordinates of $\eta NNN$ four body system.
}
\label{fig:jaco4}
\end{figure}
In this work, we investigate the $\eta^3$He system via solving the four body $\eta NNN$ Schr\"odinger equation.
The Hamiltonian is written as follows:
\begin{equation}
H=T_{\eta}+\sum_{i=1}^{3}T_{N_i}+\sum_{i<j=1}^{3}V_{N_iN_j}+\sum_{i=1}^{3}V_{\eta N_i}, \label{eq:hamil}
\end{equation}
where $T_{\eta}$ and $T_{N_i}$ are the kinetic operators of $\eta$ meson and nucleons, respectively.
$V_{N_iN_j}$ represents the interactions between nucleons and $V_{\eta N_i}$ denotes $\eta N$ potential.
In order to solve the $\eta NNN$ four body Schr\"odinger
equations, We apply the Gaussian expansion method (GEM)~\cite{hiyama2003gem,Hiyama2004} and write the four body $\eta NNN$ wave function with the $J^P=1/2^+$ and isospin $(T,T_z)=(1/2,1/2)$ as:
\begin{equation}
\begin{aligned}
\Psi_{JMTT_z}\left(\eta NNN\right)&= \sum_{c=1}^{4} \sum_{l,L}\sum_{s,S,t} \sum_{n_1,l_1}\sum_{n_2,l_2}\sum_{n_3,l_3} C_{
 \beta}^{(c)}\mathcal{A} \\
&\times\left\{\left\{\left[\left(\phi_{n_1 l_1}^{(c)}\left(\boldsymbol{r}_{c}\right) \psi_{n_2 l_2}^{(\mathrm{c})}\left(\boldsymbol{R}_{c}\right)\right)_{l} \varphi_{n_3 l_3}^{(c)}\left(\boldsymbol{\rho}_{c}\right)\right]_{L}\right.\right. \\
&\left.\times\left[\left(\chi^{1}_{1/2}\chi^{2}_{1/2}\right)_s\chi^{3}_{1/2}\right]_S\right\}_{JM} \\
&\left.\times\left[\left(\tau^{1}_{1/2}\tau^{2}_{1/2}\right)_t\tau^{3}_{1/2}\right]_{TT_z}\right\}
, \label{eq:wf1}
\end{aligned}
\end{equation}
where the sets of Jacobi coordinates ($c=1-4$) are shown in Fig.~\ref{fig:jaco4}.
Here $\beta$ denotes $\{I,L,s,S,t,n_1,l_1,n_2,l_2,n_3,l_3\}$.
$\chi$ and $\tau$ represent the spin and isospin wave function of the nucleons, respectively. 
It should be noted that both the intrinsic spin and isospin of the $\eta$ meson is 0 thus we neglect its spin and isospin wave function.
The $\mathcal{A}$ is the anti-symmetrization operator between nucleons.
The relative spatial wave functions between the nucleons and eta meson in $\eta NNN$, corresponding to the three Jacobi coordinates, $\phi_{n_1 l_1}(\boldsymbol{r})$, $\psi_{n_2 l_2}(\boldsymbol{R})$ and $\varphi_{n_3 l_3}(\boldsymbol{\rho})$, are expanded by using the following Gaussian basis functions, applying the GEM,
\begin{eqnarray}
&\phi_{n_1 l_1}({r})=r^{l_1} e^{-(r/r_{n_1})^2}Y_{l_1 m_1}(\hat{r}), \nonumber \\
&\psi_{n_2 l_2}({R})=R^{l_2} e^{-(R/R_{n_2})^2}Y_{l_2 m_2}(\hat{R}), \nonumber \\
&\varphi_{n_3 l_3}({\rho})=\rho^{l_3} e^{-(\rho/\rho_{n_3})^2}Y_{l_3 m_3}(\hat{\rho}).
\end{eqnarray}
The Gaussian variational parameters are chosen to have geometric progression below,
\begin{eqnarray}
&r_{n_1}=r_{\rm min} A_1^{n_1-1},  \quad (n_1=1 \sim n_1^{\rm max}), \nonumber \\
&R_{n_2}=R_{\rm min} A_2^{n_2-1},  \quad (n_2=1 \sim n_2^{\rm max}), \nonumber \\
&\rho_{n_3}=\rho_{\rm min} A_3^{n_3-1},  \quad (n_3=1 \sim n_3^{\rm max}). \label{eq:4}
\end{eqnarray}
Then, the eigen energies and the coefficients $C_{\gamma}$ and $C_{\beta}$ are obtained with applying the Rayleigh-Ritz variational method.
We use the AV8' $NN$ interaction which is a modified version of AV18 interaction \cite{Pudliner1997} where a tensor $NN$ interaction is included. The calculated 
binding energy of deuteron, $^3$He and $^3$H with AV8' $NN$ interaction are 2.24, 7.11 and 7.82 MeV, respectively.


\subsection{Construction of effective $\eta N$ potential}\label{sec:22}
We construct the effective $\eta N$ potential through the following equation which gives its relation with the 
scattering amplitude $F$ or the scattering length $a$:
\begin{equation}
\label{eq:kcot}
kcot{\delta}=ik+F^{-1}=\frac{1}{a}+\frac{1}{2}r_0k^2+...\;,  
\end{equation}
where $\delta$ and $r_0$ represent for the scattering phase shift and effective range, respectively.
$k$ is the wave number with $k=\sqrt{2m_{\eta N}E}$ where $E=\sqrt{s}-\sqrt{s_{th}}$.
For later convenience, we replace $E$ with $\delta\sqrt{s}$ for distinguishing with $\eta^3$He total binding energy $E$.
$m_{\eta N}$ is the reduced mass of the $\eta$ meson and nucleon.
Note that we have $a=F$ when $k\rightarrow0$.
For the further convenience, we only focus on the absolute value of the $a$ ($F$).
\footnote{Obviously, we only consider the attractive $\eta N$ potential while it is not strong enough to form a bound $\eta N$ two
body bound state. Thus, the scattering length always keeps in the same sign.}

We use Gaussian-type $\eta N$ potential as:
\begin{equation}
V_{\eta N}(r_{\eta N})=V_0(\mu/\pi)^{3/2}(\hbar c)^2/m_{\eta N}e^{-\mu r^2_{\eta N}},
\label{eq:VetaN}
\end{equation}
where the range parameter $\mu$ and potential strength $V_0$ need to be determined.
In the case of energy dependent $\eta N$ potential, 
the $V_0$ shall has the formula as $V_0(\delta\sqrt{s})$ which is dependent on the $\eta N$ center-of-mass energy $\delta\sqrt{s}$.

We then solve the two body $s$-wave $\eta N$ Schr\"odinger equation:
\begin{equation}
u^{''}(r)+k^2u(r)=2m_{\eta N}V_{\eta N}(r) \label{eq:etaNsch}
\end{equation}
under the boundary condition:
\begin{equation}
u(0)=0,\qquad u(r)\xrightarrow{r\rightarrow\infty}sin(kr)+Fke^{ikr}. \label{eq:boundary}
\end{equation}
As mentioned in the first section, in this work, we both construct the energy dependent and energy independent $\eta N$ potentials.

\begin{figure}[htbp]
\setlength{\abovecaptionskip}{0.cm}
\setlength{\belowcaptionskip}{-0.cm}
\centering
\includegraphics[width=0.44\textwidth]{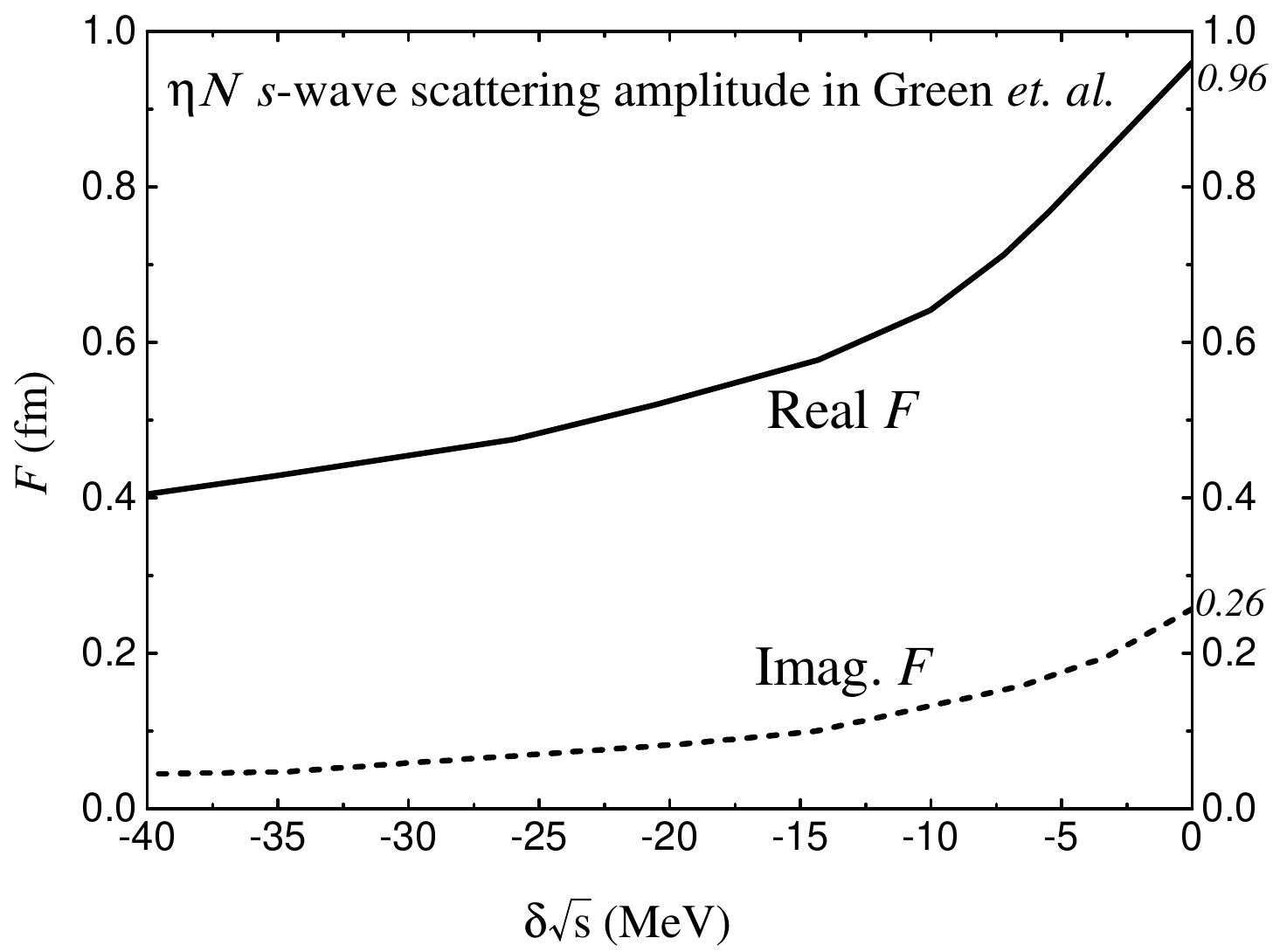}
\caption{The scattering amplitude $F(\delta\sqrt{s})$ (real part and imaginary part) at the subthreshold energy region obtained in the GW model~\cite{Green2005}.
}
\label{fig:gw}
\end{figure}

\begin{figure}[!h]
\setlength{\abovecaptionskip}{0.cm}
\setlength{\belowcaptionskip}{-0.cm}
\centering
\subfigure{
\includegraphics[width=0.45\textwidth]{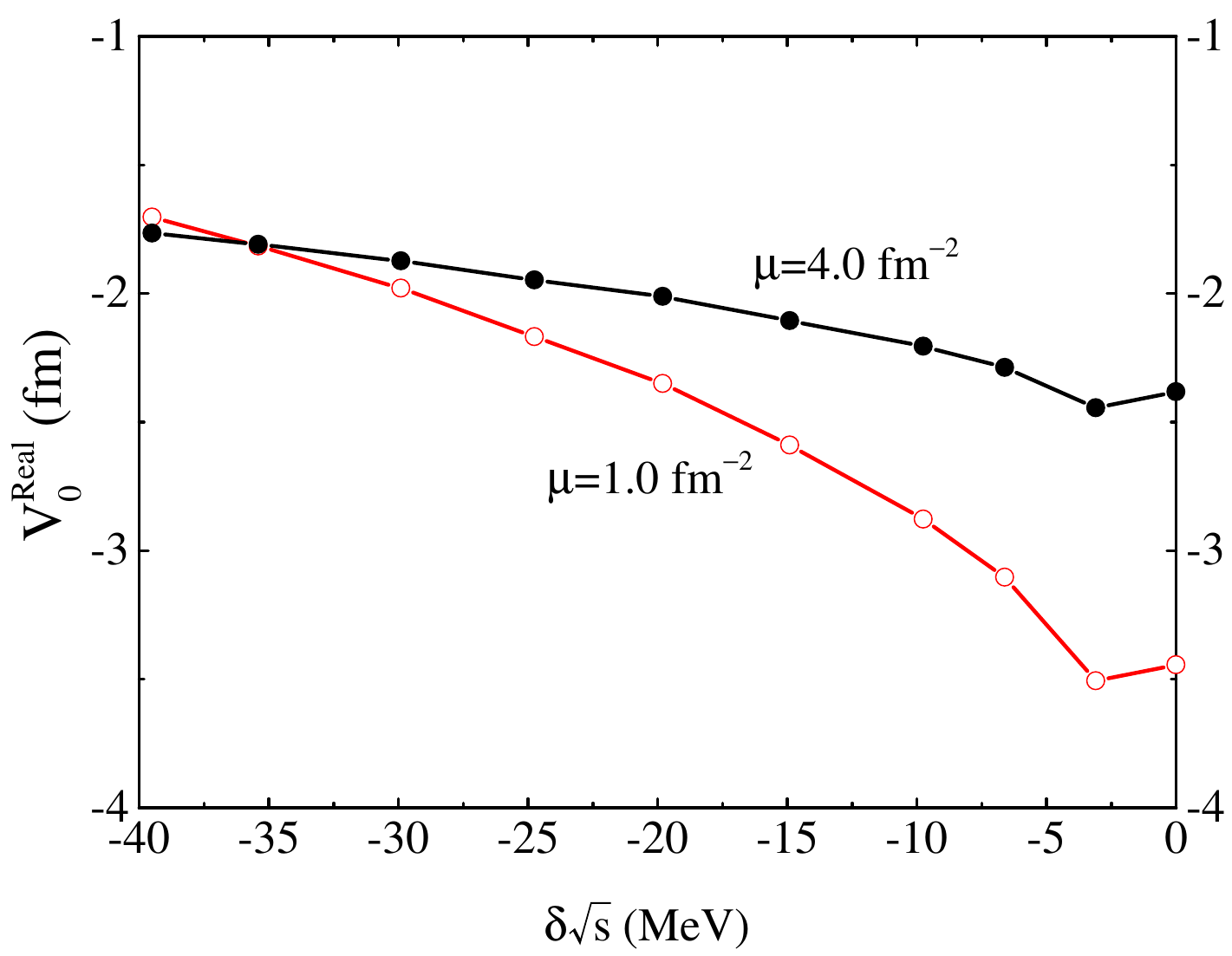}}
\hspace{1in}
\subfigure{
\includegraphics[width=0.45\textwidth]{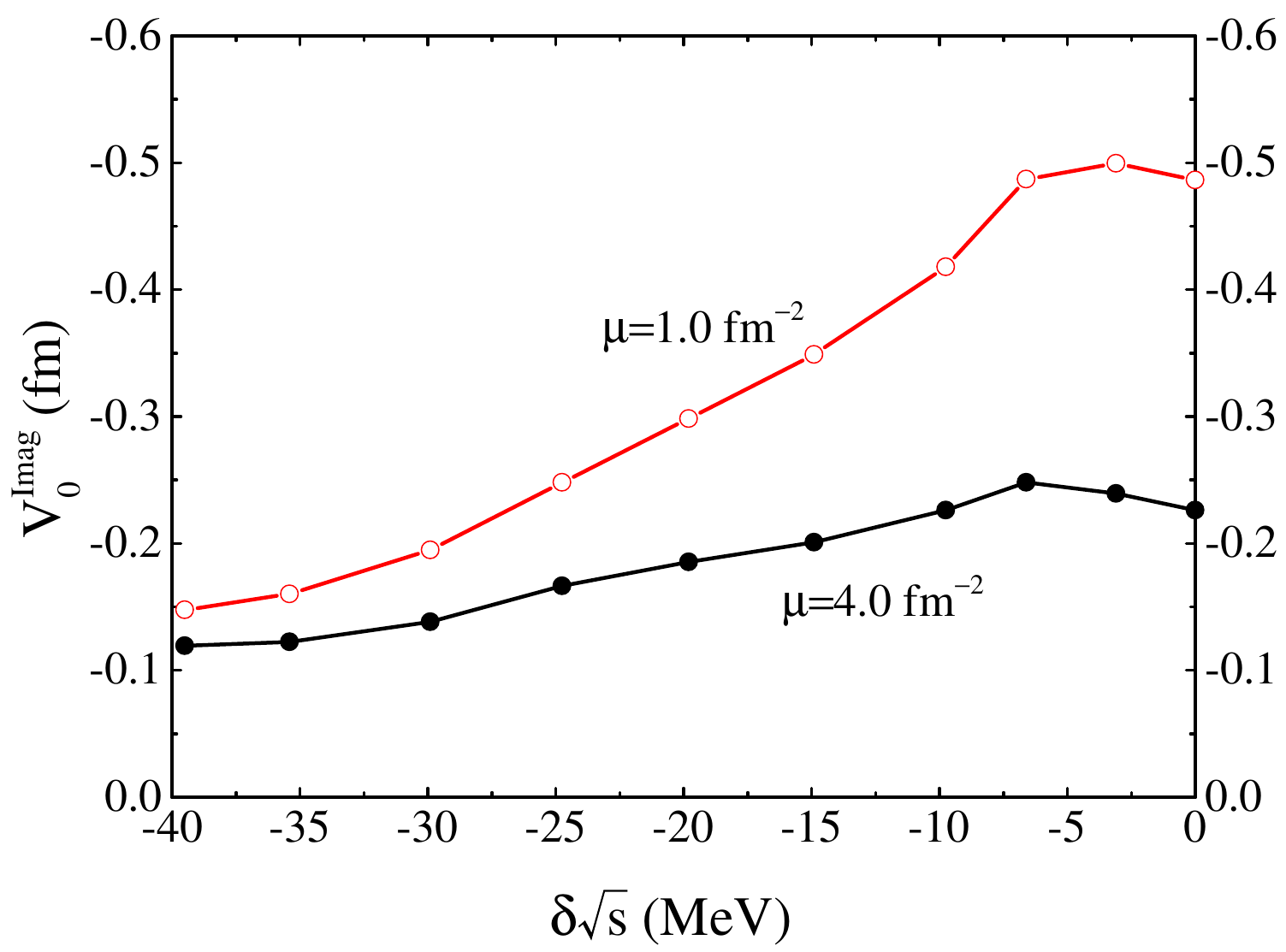}}
\caption{The strength of the $\eta N$ energy dependent potential, $V_0^{\rm Real}$ (for the in real part, up panel) and  
$V_0^{\rm Imag}$ (for the imaginary part, down panel) at the subthreshold energy region ($\delta\sqrt{s}<0$).
Two different Gaussian type potentials with range parameter $\mu$=1.0 and 4.0 fm$^{-2}$ are shown.}
\label{fig:vb-gw} 
\end{figure}

First, we construct the  energy dependent $\eta N$ potential
via reproducing the scattering amplitude $F(\delta\sqrt{s})$ given in the GW model in Ref.~\cite{Green2005} (shown in Fig.~\ref{fig:gw}).
The reason we choose the GW model is as follows:
As will be discussed in Sec.~\ref{sec:32}, the smallest scattering length to form a bound $\eta^3$He nucleus
is around $0.7\sim0.8$ fm. And as will be shown in Fig.~\ref{fig:vb-gw}, the strength of the $\eta N$ potential drops significantly as the energy goes deeper into the subthreshold energy region.
Thus, we need a relative larger value of the Real scattering length ($F(\delta\sqrt{s}=0)$) to form a bound $\eta^3$He nucleus
when the $\delta\sqrt{s}$ goes deeper. 
Among several models, the GW model gives a very large real scattering length at 0.96 fm. 
For instance, in Ref.~\cite{Cieply2013}, it gives the real scattering length at 0.67 fm which not large enough to form a bound
$\eta^3$He nucleus.

The calculated $V_0(\delta\sqrt{s})$ (real and imaginary parts) are expressed in Fig.~\ref{fig:vb-gw}.
Two different Gaussian range parameters $\mu$ (1.0 and 4.0 fm$^{-2}$) are used and the reasons for adopting these two values will be given in the end of this subsection. As we can see, both the real and imaginary parts of the $V_0(\delta\sqrt{s})$ experience a slight increase and then drops rapidly as the energy goes smaller.  

\begin{figure}[!h]
\setlength{\abovecaptionskip}{0.cm}
\setlength{\belowcaptionskip}{-0.cm}
\centering
\subfigure{
\includegraphics[width=0.45\textwidth]{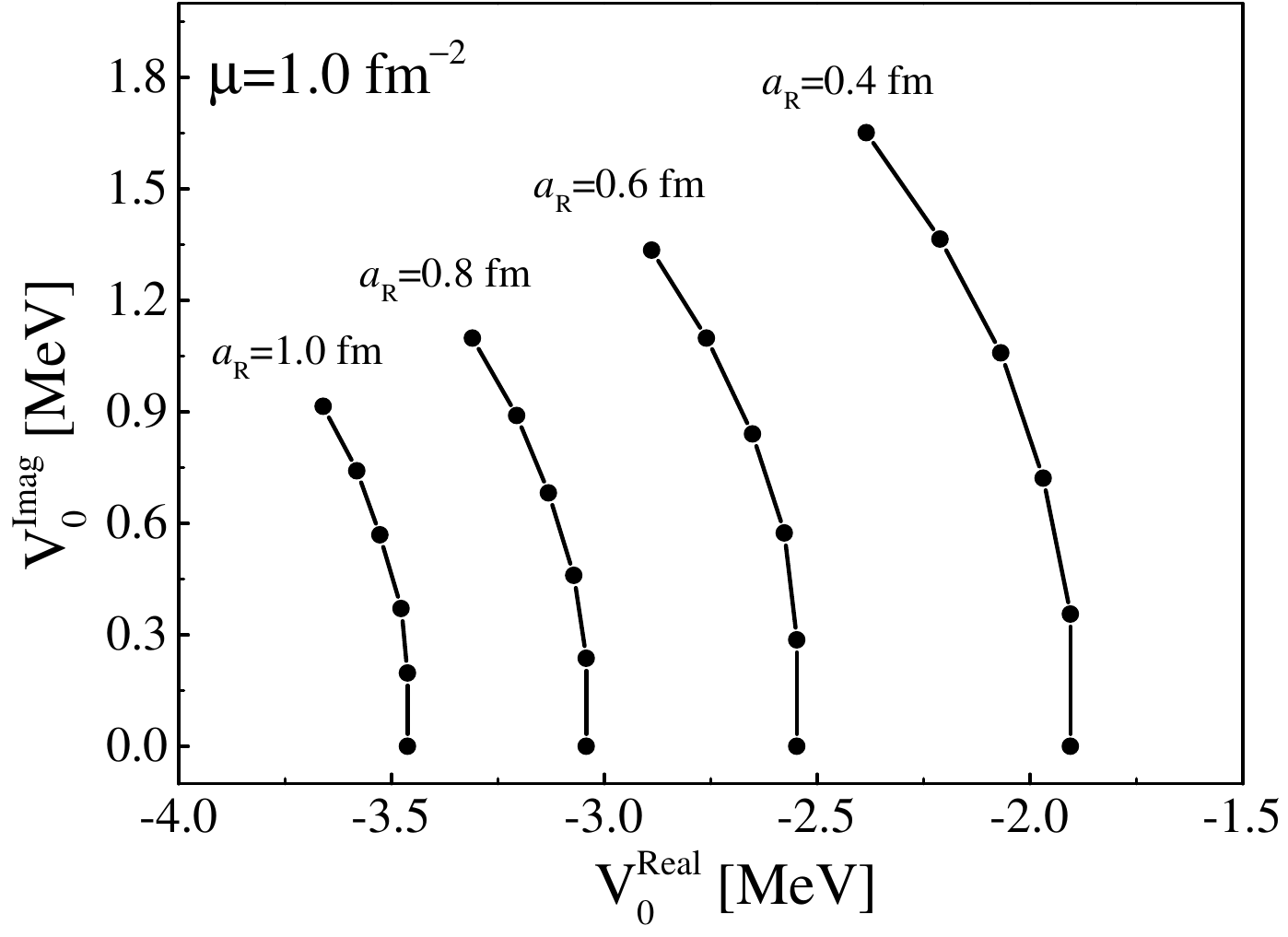}}
\hspace{1in}
\subfigure{
\includegraphics[width=0.45\textwidth]{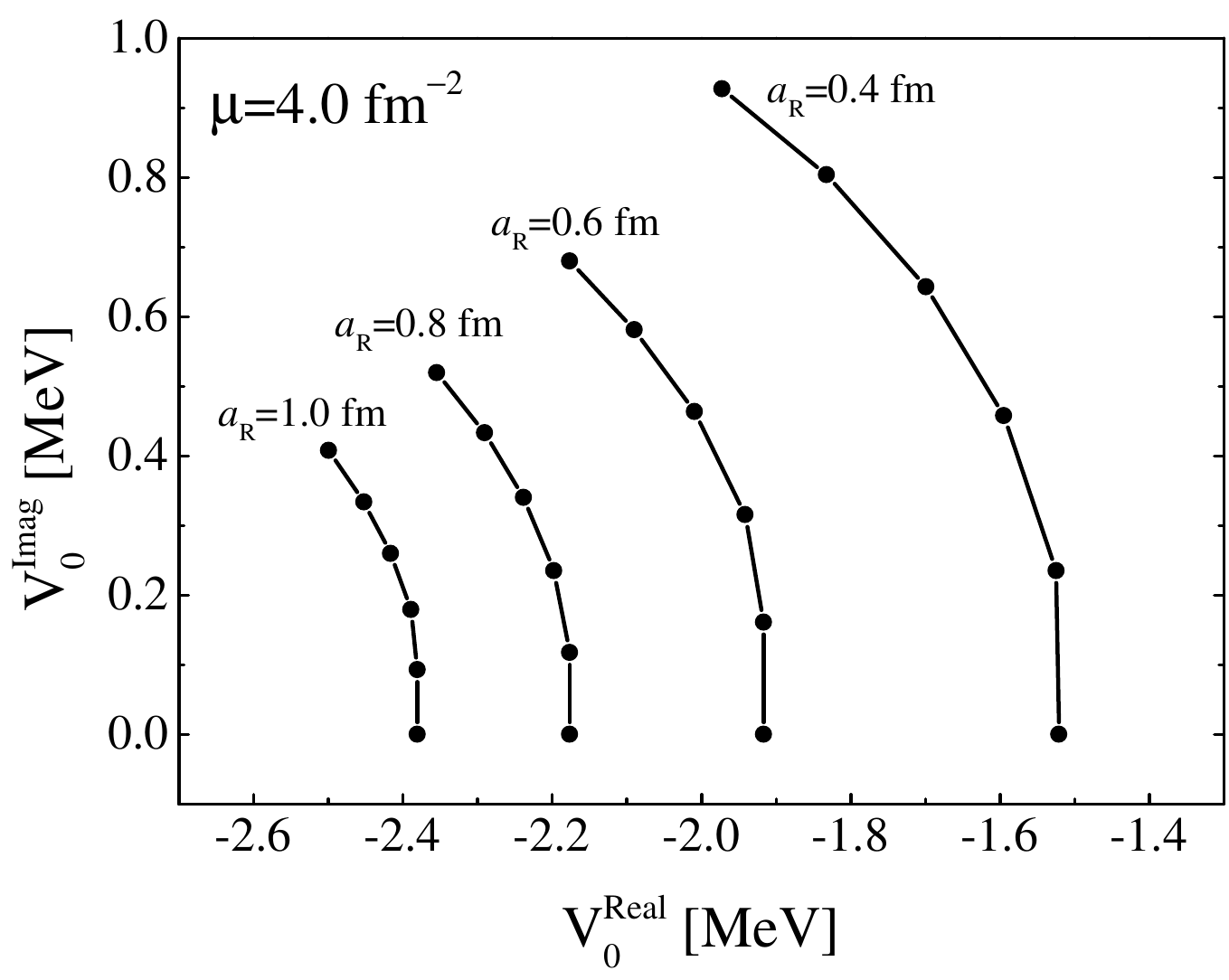}}
\caption{The complex strengths ($V_0^{\rm Real}$ and $V_0^{\rm Imag}$) of the energy independent $\eta N$ potentials under different complex scattering lengths
in the case of potential range parameter $\mu=1.0$ fm$^{-2}$ (up panel) and 4.0 fm$^{-2}$ (down panel).
In each line, the real scattering length ($a_R$) is fixed and the six points in each line are corresponding to different
imaginary scattering lengths, 0, 0.1, 0.2, 0.3, 0.4, and 0.5 fm, respectively.
}
\label{fig:scl} 
\end{figure}

Second, we build several sets of energy independent $s$-wave $\eta N$ potentials which
give the $\eta N$ scattering length within the following range as mentioned in the Sec.~\ref{sec:1}:
\begin{equation}
a_R\rightarrow 0.2\sim1.0\;{\rm fm},\;a_I\rightarrow 0.2\sim0.5\;{\rm fm}. 
\label{eq:1}
\end{equation}
Here $a_R$ and $a_I$ represent the real and imaginary $\eta N$ scattering length for short, respectively.
In Fig.~\ref{fig:scl}, we depict the $V_0$ ($V_0^{\rm Real}$ for the real part and $V_0^{\rm Imag}$ for the imaginary part) 
by separating them with different scattering length.
Namely, in Fig.~\ref{fig:scl}, each point has a corresponding scattering length ($a_R$, $a_I$).
In each folding line, the $a_R$ is fixed and the six points represent the six different $a_I$,
which are $0\sim0.5$ fm (step = 0.1 fm) from down to the top.

One interesting thing in Fig.~\ref{fig:scl} is that, the $V_0^{\rm Real}$ has no change when the $a_I$ grows from 0 to 0.1 fm
but it has more and more significant enhancement when the $a_I$ is around 0.5 fm. Namely, each folding line turns flatter as
the $a_I$ grows, specially in the case of $\mu=4.0$ fm$^{-2}$.

\begin{table}[h]
\caption{The momentum scale parameter $\Lambda$ used in several EFT models.}
\begin{tabular}{p{2cm}<{\centering}p{2cm}<{\centering}p{2cm}<{\centering}}
  \hline\hline
    \noalign{\vskip 0.1 true cm}
     Refs & \cite{plb10,Kaiser1995} & \cite{plb11}        \\
       \noalign{\vskip 0.05 true cm}
  \hline
    \noalign{\vskip 0.05 true cm}
$\Lambda$ (fm$^{-1}$) & 3.9 &3.2 \\
  \noalign{\vskip 0.1 true cm}
 \hline
 \hline
\end{tabular}
\label{tab:scale}
\end{table}

It is obvious that our potentials strongly depend on the range parameter $\mu$. It is often identified with
the momentum cutoff $\Lambda$ ($\Lambda=2\sqrt{\mu}$) which is used to treat the divergent loop 
integrals in on-shell EFT $N^*$(1535) models \cite{Mai2012,Inoue2002}. In Table \ref{tab:scale}, we give the $\Lambda$ values used in several different EFT $N^*$(1535) models. It gives a range of $\mu$ around $2\sim4$ fm$^{-2}$. It should be noted that in Ref.~\cite{Cieply2013}, $\Lambda=6.6$ fm$^{-1}$ is used which gives the potential range $1/\sqrt{\mu}=0.3$ fm.
But according to the Ref.~\cite{gal2015plb}, choosing a potential range smaller than 0.47 fm ($\mu\sim4$ fm$^{-2}$) would be inconsistent with staying within a purely hadronic basis. Therefore, we won't consider any $\mu$ values which are larger than 4 fm$^{-2}$ and use $\mu=1.0$ and $4.0$ fm$^{-2}$ as a benchmark in this work.

\section{Results}
\subsection{Energy dependent $\eta N$ potential}\label{sec:31}
In this subsection, we show the numerical results of the $\eta^3$He nucleus using the energy dependent $\eta N$ potential in Fig.~\ref{fig:vb-gw}.
Similar as Ref.~\cite{gal2015plb}, the $\eta N$ center-of-mass energy $\delta\sqrt{s}$ in the environment of the $\eta^3$He nucleus
can be given with:
\begin{equation}
\begin{aligned}
\left\langle\delta \sqrt{s}\right\rangle&=-\frac{E}{3}+\frac{2B_\eta}{3}-\beta_N \frac{1}{3}\left\langle T_N\right\rangle-\beta_\eta\left(\frac{2}{3}\right)^2\left\langle T_\eta\right\rangle
\label{eq:deltas}
\end{aligned}
\end{equation}
where $\beta_{N(\eta)}= m_{N(\eta)}/(m_N+m_\eta)$, $T_N$ and $T_\eta$ are the nuclear and $\eta$ kinetic energy operators evaluated in center-of-mass frame. $E$ is the total binding energy of the $\eta^3$He nucleus and $B_\eta$ is the $\eta$ separation energy with respect to the $^3$He threshold.
\begin{figure}[htbp]
\setlength{\abovecaptionskip}{0.cm}
\setlength{\belowcaptionskip}{-0.cm}
\centering
\includegraphics[width=0.46\textwidth]{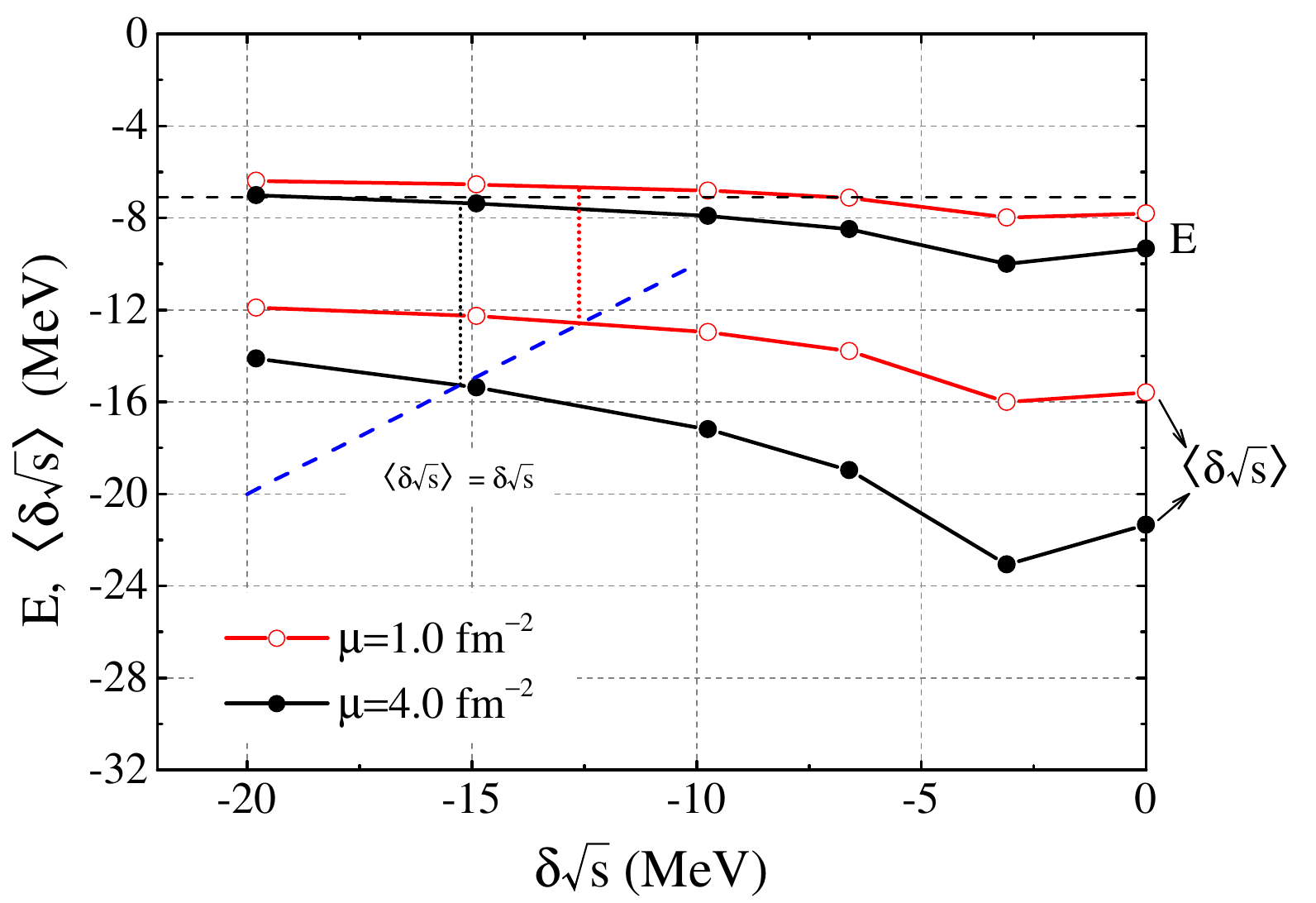}
\caption{The relation between the $\eta N$ center-of-mass energy $\delta \sqrt{s}$ and total binding energy $E$ in the $\eta^3$He system. The relation between the calculated $\langle\delta \sqrt{s}\rangle$ (Eq.~\ref{eq:deltas}) and the $\delta\sqrt{s}$ is also given. The $\mu=1.0$ fm$^{-2}$ case and the $\mu=4.0$ fm$^{-2}$ case are shown with red hollow circle and black solid circle, respectively.
The blue dashed line is drawn with $\langle\delta \sqrt{s}\rangle=\delta\sqrt{s}$.}
\label{fig:etahe3-gw}
\end{figure}

Then, the binding energy of the $\eta^3$He nucleus is obtained with a self-consistent procedure.
In Fig.~\ref{fig:vb-gw}, we give the relation between the $\eta N$ potential and the $\delta\sqrt{s}$.
With applying the $\eta N$ potential into the four body calculation, we obtain the relation between the 
binding energy $E$ ($B_\eta$) and the $\delta\sqrt{s}$.
Then, based on Eq.~\ref{eq:deltas}, we obtain the relation between $\langle\delta \sqrt{s}\rangle$ and $\delta\sqrt{s}$.
The relations between the $E$ and the $\delta\sqrt{s}$ and the relation between $\langle\delta \sqrt{s}\rangle$ and $\delta\sqrt{s}$ are shown in Fig.~\ref{fig:etahe3-gw}. The blue dashed curve ($\langle\delta \sqrt{s}\rangle=\delta\sqrt{s}$) certainly give the self consistent value of the $\eta^3$He binding energy $E$. Note that here we temporarily only consider the real part of the $\eta N$ potential and treat the imaginary $\eta N$ potential perturbatively.

In the case of $\mu=1.0$ fm$^{-2}$, it gives no self-consistent bound state of $\eta^3$He. As shown in Fig.~\ref{fig:etahe3-gw}, the
intersection point of the line $E(\delta\sqrt{s})$ and the line $\langle\delta \sqrt{s}\rangle=\delta\sqrt{s}$ is above the $^3$He threshold. 

As for $\mu=4.0$ fm$^{-2}$ case, the intersection point is below the $^3$He threshold giving a bound state.
The $\eta$ binding energy ($B_\eta$), $\langle\delta \sqrt{s}\rangle$ and decay width $\Gamma$ are shown in Table ~\ref{tab:tab-compare}.
For comparison, three similar calculations in Refs.~\cite{gal2015plb,gal2017plb-pion} are shown in Table \ref{tab:tab-compare}, which also employ the energy dependent $\eta N$ potential reproducing the GW scattering amplitude in Fig.~\ref{fig:gw} but with a different few-body method and different $NN$ potentials. 
All of these works are consistent with each other giving a weakly bound $\eta^3$He nucleus and the decay width around 1.5 MeV\footnote{The decay widths given in Ref.~\cite{gal2015plb} are corrected in Ref.~\cite{gal2017plb-pion} but they didn't give a definite value. But the authors indicated that the decay widths should be similar as the one in Ref.~\cite{gal2017plb-pion}}.

\begin{table}[h]
\caption{The $\eta$ separation energy $B_\eta$ (MeV), the decay width $\Gamma$ (MeV) and the $\langle\delta \sqrt{s}\rangle$ (MeV)
obtained in the case of $\mu=4.0$ fm$^{-2}$ are in the second row. Three similar works taken from Refs.~\cite{gal2015plb,gal2017plb-pion} are shown in row 3 to 5, which also use the energy dependent $\eta N$ potential but with a different few-body method. 
}
\begin{tabular}{p{2.5cm}<{\centering}p{1.0cm}<{\centering}p{1.0cm}<{\centering}p{1.0cm}<{\centering}}
  \hline\hline
   \noalign{\vskip 0.1 true cm}
   & $\langle\delta\sqrt{s}\rangle$ & $B_{\eta}$ &  $\Gamma$\\
   \noalign{\vskip 0.1 true cm}
  \hline
   \noalign{\vskip 0.1 true cm}
  AV8'(present) & -15.25 & 0.19 & 1.71 \\
    \noalign{\vskip 0.1 true cm}
 \hline
   \noalign{\vskip 0.1 true cm}
  AV4' (Ref.~\cite{gal2015plb}) &-15.83 & 0.04 & -- \\
 \hline
  \noalign{\vskip 0.1 true cm}
  MN (Ref.~\cite{gal2015plb}) &  -13.52 & 0.95 & -- \\
  \hline
  \noalign{\vskip 0.1 true cm}
  Ref.~\cite{gal2017plb-pion} &  -21.30 & 0.30 & 1.46 \\
  \hline
   \hline
\end{tabular}
\label{tab:tab-compare}
\end{table}

\subsection{Energy independent $\eta N$ potential}\label{sec:32}
As shown in Table \ref{tab:tab-compare}, the decay width given by several few body calculations including the present work which reproduce the GW scattering amplitude are all between $1\sim2$ MeV,
which are much less than the observed values in Ref.~\cite{Adlarson2019plb} of $>$20 MeV and in Ref.~\cite{Adlarson2019prc} of 5$\sim$50 MeV. 
As we mentioned before, the strength of the energy dependent $\eta N$ potential decreases as it goes deeper into the subthreshold energy region.
Thus, in order to obtain a larger decay width, we exclude this suppression and 
employ the $\eta^3$He system with the energy independent $\eta N$ potential.

\begin{figure}[htbp]
\setlength{\abovecaptionskip}{0.cm}
\setlength{\belowcaptionskip}{-0.cm}
\centering
\includegraphics[width=0.46\textwidth]{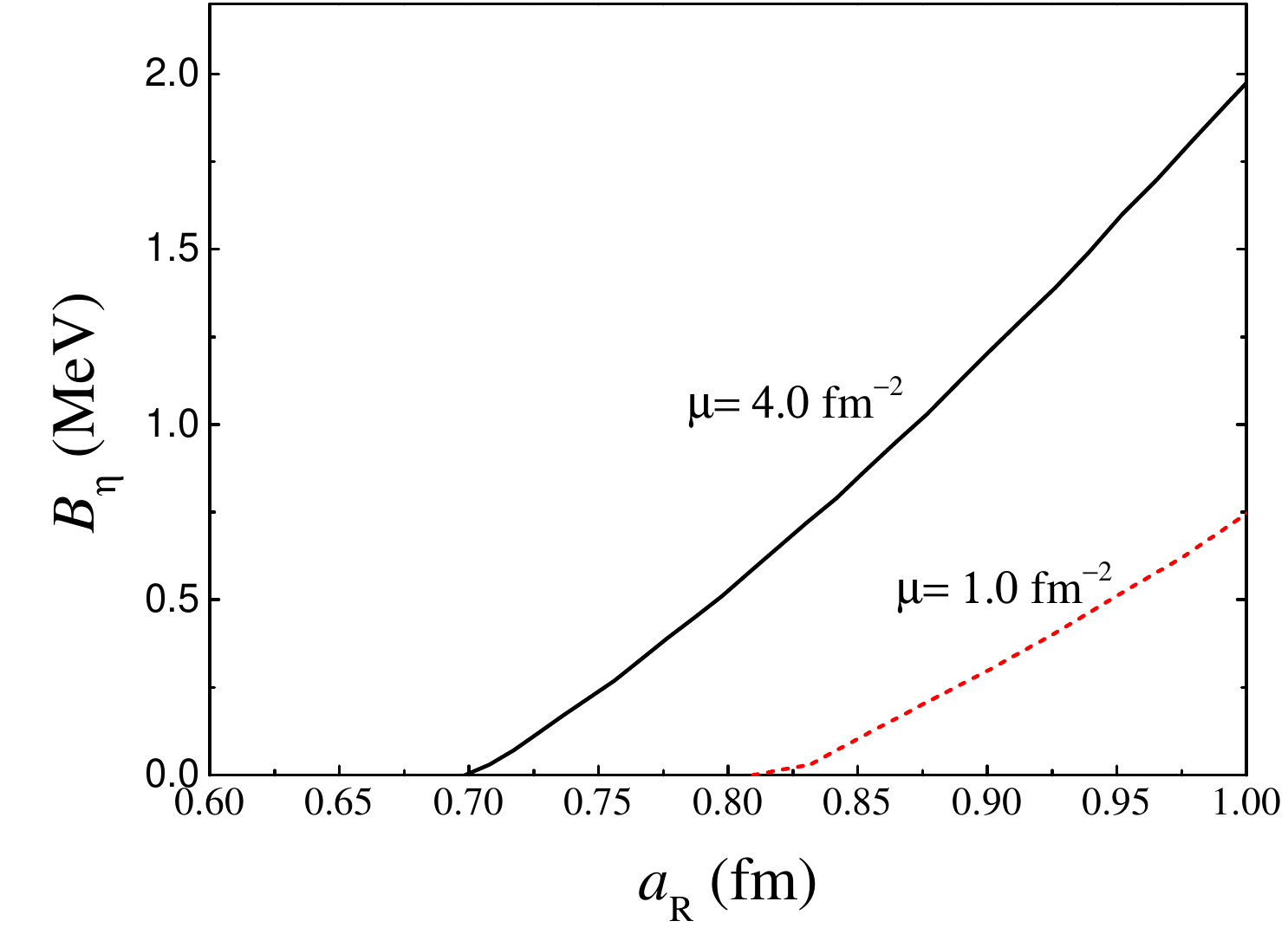}
\caption{The relations between the $B_\eta$ values in the $\eta^3$He system and the $\eta N$ scattering lengths (only real part). The black solid line represents the $\mu=4.0$ fm$^{-2}$ and the red dashed line represents the $\mu=1.0$ fm$^{-2}$.}
\label{fig:etahe3-real}
\end{figure}

First, we neglect the imaginary part of the $\eta N$ potential.
In Fig.~\ref{fig:etahe3-real}, we show the $B_\eta$ under different $a_R$.
We find that to form a $\eta^3$He bound system, the smallest scattering length for
range parameter $\mu=4.0$ fm$^{-2}$ is around 0.7 fm. And the smallest scattering length for $\mu=1.0$ fm$^{-2}$ is 0.83 fm.
The corresponding binding energy with respect to the $^3$He threshold is between $0\sim2$ MeV.

Then, we include the imaginary $\eta N$ potential and solve the four body complex Schr\"odinger equation.
Note that here we do not treat the imaginary $\eta N$ potential perturbatively but fully diagonalize the Hamiltonian.
We search the $\eta^3$He bound state of each set of the $\eta N$ potentials in Fig.~\ref{fig:scl}. 
As we mentioned before, the smallest scattering lengths are 0.7 and 0.83 fm in order to form a $\eta^3$He bound state, respectively for $\mu=4.0$ and $\mu=1.0$ fm$^{-2}$ cases when we only consider the real $\eta N$ potential. 
And considering the fact that the imaginary $\eta N$ potential have negative contribution to form the bound nuclei, it is not necessary
to consider the cases when $a_R<0.83$ fm of $\mu=1.0$ fm$^{-2}$ case and $a_R<0.70$ fm of $\mu=4.0$ fm$^{-2}$ case.

\begin{figure}[htp]
\setlength{\abovecaptionskip}{0.cm}
\setlength{\belowcaptionskip}{-0.cm}
\centering
\subfigure{
\includegraphics[width=0.41\textwidth]{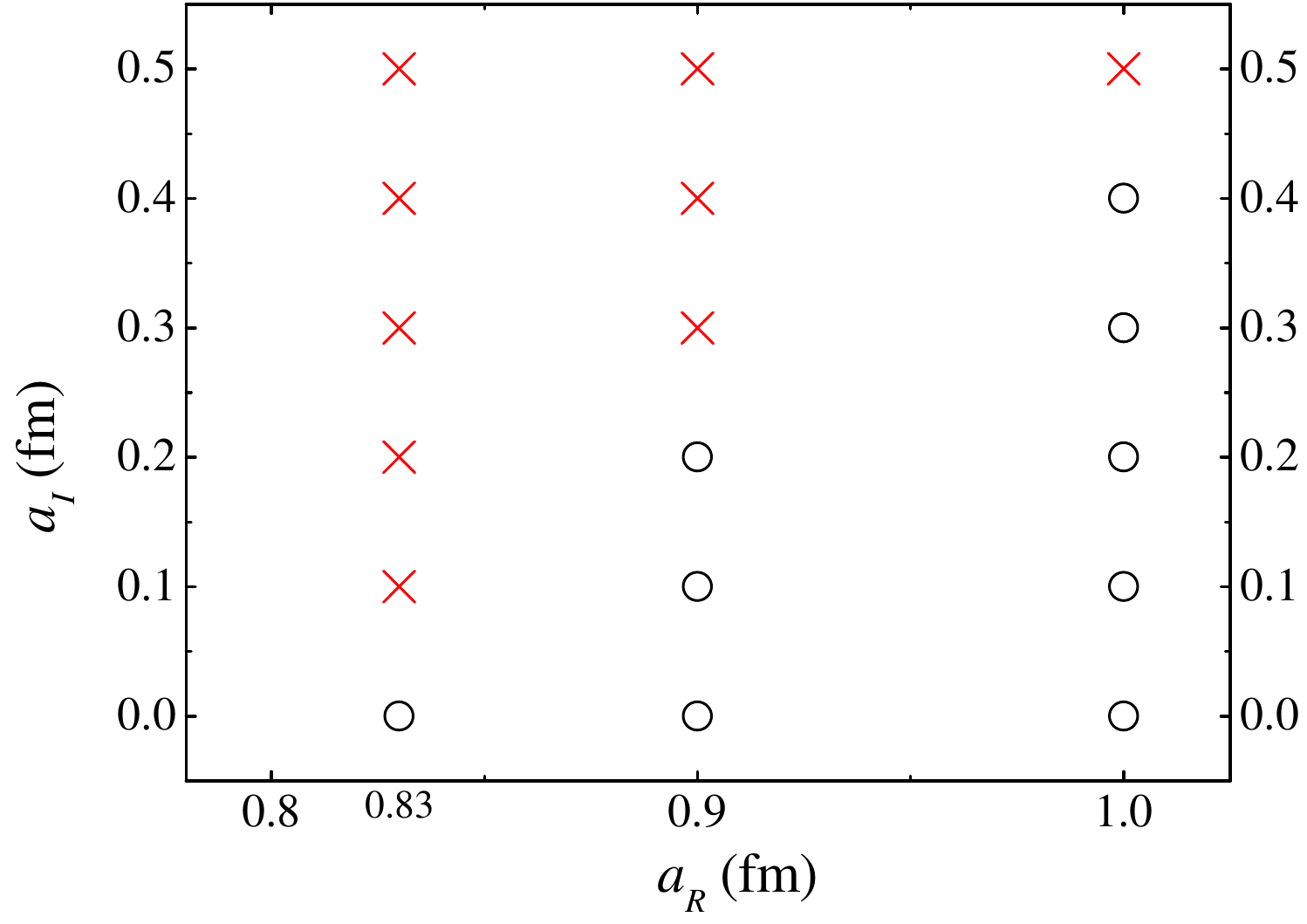}}
\hspace{1in}
\subfigure{
\includegraphics[width=0.41\textwidth]{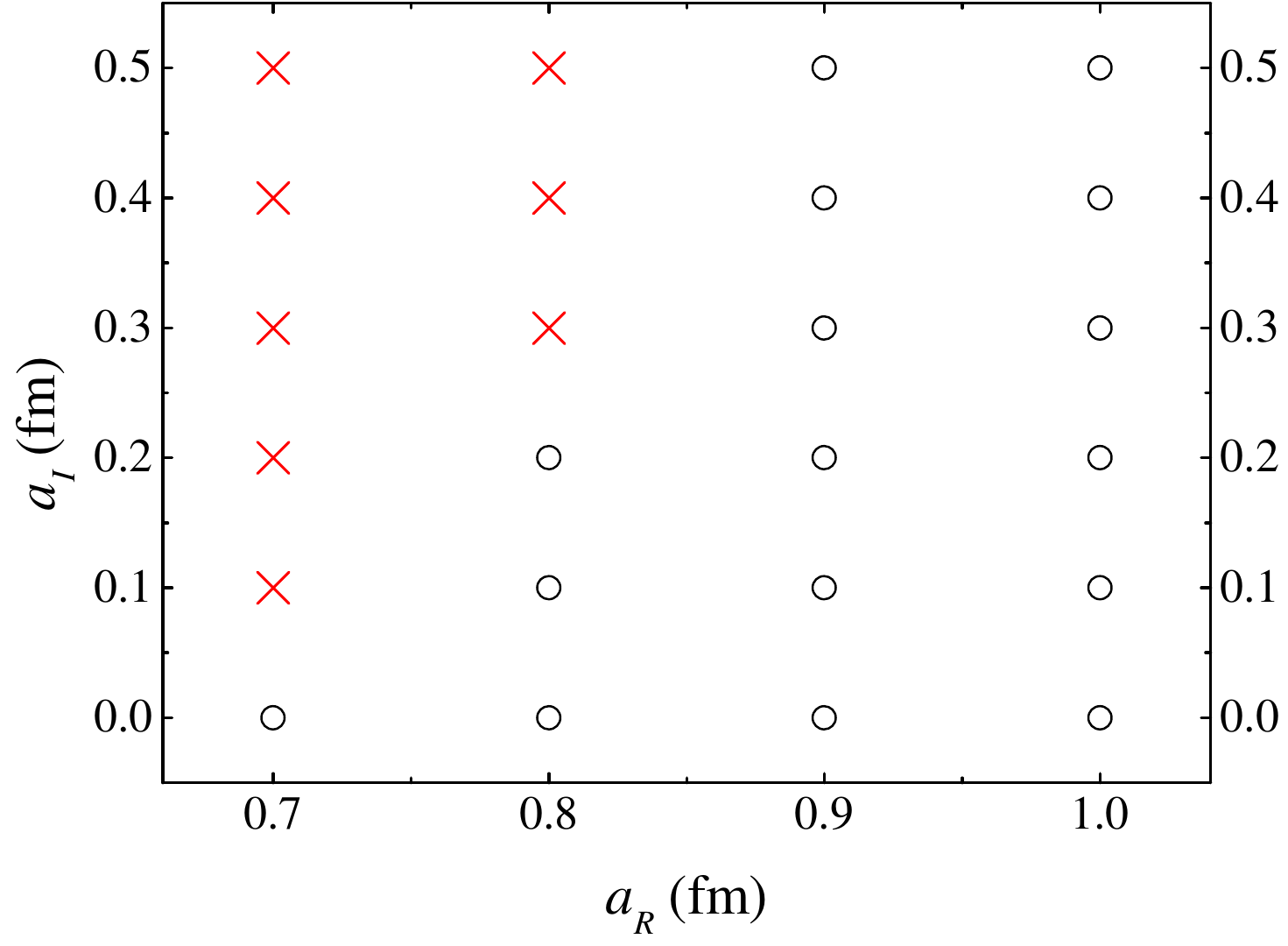}}
\caption{Existences of the bound $\eta^3$He system under different complex scattering lengths ($a_R,a_I$) when the range parameter $\mu=1.0$ fm$^{-2}$ (up panel) and 4.0 fm$^{-2}$ (down panel). The circles mean there exist a bound state while the crosses mean there don't.}
\label{fig:cross} 
\end{figure}

In Fig.~\ref{fig:cross}, we show whether there exists a bound $\eta^3$He nucleus under different scattering lengths, ($a_R,a_I$).
The black circles mean there exists a bound $\eta^3$He nucleus while the 
crosses mean there doesn't exist any.
It illustrates how the imaginary scattering length gives a negative contribution in forming a bound $\eta^3$He nucleus. 
When $a_R$ reduces from 1.0 to 0.83 fm, the cases which have a bound state also reduce. 
Similar behaviors occur in the case of $\mu=4.0$ fm$^{-2}$.
The difference is that when $a_R$ is around 0.9 to 1.0 fm, the $\eta^3$He nucleus is always bound due to the relative larger binding energy when $a_I=0$. Thus, it keeps bound when $a_I$ increases.
 
Besides, in Table \ref{tab:E-etahe3-1} and Table \ref{tab:E-etahe3-4}, we show the $B_\eta$ and $\Gamma$ values of some of the black circles in Fig.~\ref{fig:cross}, in which the 
$\eta^3$He nucleus is bound. 
It should be noted that the decay widths grow significantly as the imaginary scattering length is around 0.5 fm, which are around 10 MeV and much larger than the $B_\eta$ value.
Thus, with the energy independent $\eta N$ potential, we may obtain a much larger decay width.

\begin{table}[tbh]
\caption{The $B_\eta$ (MeV) and decay width $\Gamma$ (MeV) of $\eta^3$He system under different complex scattering length ($a_R,a_I$) (in unit of fm).
The potential range parameter $\mu$ is $1.0$ fm$^{-2}$.}
\begin{tabular}{p{2.5cm}<{\centering}p{2.5cm}<{\centering}}
  \hline\hline
  \noalign{\vskip 0.1 true cm}
    $(a_R,\;a_I)$ & $(B_\eta,\;\Gamma)$         \\
    \noalign{\vskip 0.1 true cm}
  \hline 
  \noalign{\vskip 0.05 true cm}
   $(1.0,0)$   & $(0.75,-)$      \\
    \hline
    \noalign{\vskip 0.05 true cm}
   $(1.0,0.2)$ & $(0.57,1.88)$  \\
    \hline
    \noalign{\vskip 0.05 true cm}
   $(1.0,0.4)$ & $(0.15,4.23)$      \\
    \hline
    \noalign{\vskip 0.05 true cm}
     $(0.9,0)$ & $(0.31,-)$      \\
    \hline
    \noalign{\vskip 0.05 true cm}
     $(0.9,0.2)$ & $(0.11,1.64)$      \\
    \hline
    \noalign{\vskip 0.05 true cm}
     $(0.83,0)$ & $(0.12,-)$      \\
    \hline
\end{tabular}
\label{tab:E-etahe3-1}
\end{table}

\begin{table}[tbh]
\caption{The $B_\eta$ (MeV) and decay width $\Gamma$ (MeV) of $\eta^3$He system under different complex scattering length ($a_R,a_I$) (in unit of fm).
The potential range parameter $\mu$ is $4.0$ fm$^{-2}$.}
\begin{tabular}{p{2.5cm}<{\centering}p{2.5cm}<{\centering}}
  \hline\hline
  \noalign{\vskip 0.1 true cm}
    $(a_R,\;a_I)$ & $(B_\eta,\;\Gamma)$         \\
    \noalign{\vskip 0.1 true cm}
  \hline 
  \noalign{\vskip 0.05 true cm}
   $(1.0,0)$   & $(2.00,-)$      \\
    \hline
    \noalign{\vskip 0.05 true cm}
   $(1.0,0.3)$ & $(1.65,5.24)$  \\
    \hline
    \noalign{\vskip 0.05 true cm}
   $(1.0,0.5)$ & $(1.46,9.72)$      \\
    \hline
    \noalign{\vskip 0.05 true cm}
     $(0.9,0)$ & $(1.25,-)$      \\
    \hline
    \noalign{\vskip 0.05 true cm}
     $(0.9,0.3)$ & $(0.68,4.81)$      \\
    \hline
    \noalign{\vskip 0.05 true cm}
     $(0.9,0.5)$ & $(0.22,8.44)$      \\
    \hline
     \noalign{\vskip 0.05 true cm}
     $(0.8,0)$ & $(0.54,-)$      \\
    \hline
     \noalign{\vskip 0.05 true cm}
     $(0.8,0.2)$ & $(0.15,2.54)$      \\
    \hline
     \noalign{\vskip 0.05 true cm}
     $(0.7,0)$ & $(0.03,-)$      \\
    \hline
\end{tabular}
\label{tab:E-etahe3-4}
\end{table}

As shown in Table \ref{tab:E-etahe3-4}, the decay width is around 10 MeV when $a_I\sim0.5$ fm.
So we think the decay width might exceed 20 MeV when $a_I\sim1.0$ fm if there still exist a bound $\eta^3$He nucleus with
$a_R\sim$1.0 fm.
Although there has been no theoretical model which gives the imaginary scattering length up to 
1 fm, we still want to have a glimpse of the binding energy and the decay width in the case of $a_I=1.0$ fm.
In Table \ref{tab:decaywidth}, the binding energy and the decay width in the case of $(a_R,a_I)=(1.0,1.0)$ (in unit of fm)
are shown together with the $(a_R,a_I)=(1.0, 0.3)$ case ($\mu=4.0$ fm$^{-2}$).
It is surprise that the binding energy, 2.71 MeV, is even larger than the ($1.0,0.3$) and ($1.0,0.5$) cases.
This may be due to the behavior we mentioned in the second last paragraph in Sec.~\ref{sec:22}:
The $V_0^{\rm Real}$ are enhanced more and more significantly as $a_I$ is larger than 0.5 fm. 
Therefore, the binding energy might have a inverse behavior and grows larger as the $a_I$ increases from 0.5 fm.

Shown in Table \ref{tab:decaywidth}, the decay width in the $(a_R,a_I)=(1.0, 1.0)$ case is around 20 MeV, just as we expect.
Together with two calculated results, we also put the experimental values in Refs~\cite{Adlarson2019plb,Adlarson2019prc}.
If the observed $\Gamma>20$ MeV in Ref.~\cite{Adlarson2019plb} is positive, we may give a explanation here.
But anyway, we need more definite experimental conclusions and values.

\begin{table}[tbh]
\caption{The binding energies ($B_{\eta}$) and the decay widths ($\Gamma$) of the $\eta^3$He nucleus obtained in the present work
in the case of scattering length $(a_R,a_I)=(1.0,0.3)$ and $(1.0,1.0)$ (in unit of fm). The range parameter $\mu=4.0$ fm$^{-2}$. 
The last two columns are the experimental values taken from Ref.~\cite{Adlarson2019plb} and Ref.~\cite{Adlarson2019prc}, respectively. 
The $B_{\eta}$ and $\Gamma$ are in unit of MeV.}
\begin{tabular}{p{0.5cm}<{\centering}p{1.3cm}<{\centering}p{1.3cm}<{\centering}p{1.1cm}<{\centering}p{1.1cm}<{\centering}}
  \hline\hline
   \noalign{\vskip 0.1 true cm}
   &(1.0, 0.3)  & (1.0, 1.0) & Ref.~\cite{Adlarson2019plb} & Ref.~\cite{Adlarson2019prc}       \\
   \noalign{\vskip 0.1 true cm}
  \hline
   \noalign{\vskip 0.1 true cm}
  $\Gamma$ & 5.24 &20.51 & $>20$ & $(5,\;50)$ \\
    \noalign{\vskip 0.1 true cm}
 \hline
   \noalign{\vskip 0.1 true cm}
  $B_{\eta}$ & 1.65 & 2.71 & $(0,\;15)$ & $(0,\;40)$ \\  
 \hline
 \hline
\end{tabular}
\label{tab:decaywidth}
\end{table}

\section{Summary}\label{sec:4}
We examine the possible existence of the $\eta^3$He mesic nucleus via solving the four body Hamiltonian with the Gaussian expansion method. We construct the effective $s$-wave energy dependent $\eta N$ potential, which reproduces the subthreshold $\eta N$ scattering amplitude given by the GW model~\cite{Green2005}. The $\eta$ separation energy and the decay width are then obtained by the self-consistent procedure~\cite{gal2017plb-pion}, with 0.19 and 1.71 MeV, respectively. These values are consistent with the previous self-consistent few-body calculations with the same energy dependent $\eta N$ potential derived from the GW model~\cite{gal2015plb,gal2017plb-pion}.

We also construct several sets of the energy independent $s$-wave $\eta N$ potentials and each set has
a corresponding complex $\eta N$ scattering length ($a_R,a_I$). Then, we give whether or not the $\eta^3$He is bound under specific scattering length. The possibility of a large decay width is then investigated though these sets of $\eta N$ potential. We find that when the complex scattering length ($a_R,a_I$) is (1.0 fm, 0.3 fm), the $\eta^3$He is bound by 1.65 MeV with the decay width of about 5 MeV. And it's bound by 1.46 MeV with the width of about 10 MeV at (1.0 fm, 0.5 fm).

\section*{Acknowledgments}
The authors thank Dr. Q. Meng for helpful discussions. We thank Prof. E. Hiyama and Prof. M. Kamimura for supports in the calculation code . 
This work is supported by the Strategic Priority Research Program of Chinese Academy of Sciences, Grant No. XDB34030301 (XC and QW), Guangdong Major Project of
Basic and Applied Basic Research, Grant No. 2020B0301030008 (XC and GX), and 
National Natural Science Foundation of
China, Grant No. 12233002 (QW).

\end{document}